# Public sentiment analysis and topic modeling regarding ChatGPT in mental health on Reddit: Negative sentiments increase over time

Yunna Cai[1], Fan Wang[1]†, Haowei Wang[1] and Qianwen Qian[1].

[1]School of Information Management, Wuhan Univerisity

## Abstract

In order to uncover users' attitudes towards ChatGPT in mental health, this study examines public opinions about ChatGPT in mental health discussions on Reddit. Researchers used the bert-base-multilingual-uncased-sentiment techniques for sentiment analysis and the BERTopic model for topic modeling. It was found that overall, negative sentiments prevail, followed by positive ones, with neutral sentiments being the least common. The prevalence of negative emotions has increased over time. Negative emotions encompass discussions on ChatGPT providing bad mental health advice, debates on machine vs. human value, the fear of AI, and concerns about Universal Basic Income (UBI). In contrast, positive emotions highlight ChatGPT's effectiveness in counseling, with mentions of keywords like "time" and "wallet." Neutral discussions center around private data concerns. These findings shed light on public attitudes toward ChatGPT in mental health, potentially contributing to the development of trustworthy AI in mental health from the public perspective.

## 1 Introduction

WARNING: This paper contains examples and descriptions which are depressive or aggressive in nature.

Since its inception, ChatGPT has been regarded as a significant opportunity for various fields, including mental health. Academia has assessed ChatGPT's outstanding performance in various mental health tasks, signaling a new era in internet-based psychological interventions (P et al., 2023). However, the application of ChatGPT in the field of mental health also raises ethical concerns (Tirth et al., 2023). It is imperative to understand the public's attitudes towards this and explore the future of generative artificial intelligence in mental health. However, there is currently a lack of empirical research to unveil users' perspectives on ChatGPT's application in mental health. Therefore, the objective of this study is to delve into the emotions and viewpoints of these users when it comes to ChatGPT in mental health.

This study poses the following questions:

1. What are the overall sentiments and topics in discussions related to ChatGPT in mental health?
2. How do the overall sentiments and topics change over time?
3. What are the topics associated with positive and negative sentiments in discussions related to ChatGPT in mental health?

To address these questions, this study seeks to examine public sentiments and opinions regarding ChatGPT in mental health from Reddit (a popular social media platform). We employed the bert-base-multilingual-uncased-sentiment model for sentiment analysis and utilized BERTopic for topic modeling. Both models exhibited exemplary performance in their designated tasks and are commonly employed in scholarly circles for sentiment and topic analysis (Maarten, 2022).

The results revealed that in discussions concerning ChatGPT in mental health, negative sentiments outweighed positive sentiments, with neutral sentiments being the least prevalent. Over time, there was a continuous increase in the proportion of negative sentiments. Users discussed various aspects, including their overall experiences with ChatGPT in mental health, prompts, associated risks (privacy and societal implications), and the impact of version updates.

These findings enrich our comprehension of the public's perceptions regarding the application of ChatGPT in mental health. Significantly, the outcomes of this study will provide valuable guidance for the development of reliable Language Models (LLMs) in the mental health domain, both for industry and government applications.

---

† Corresponding author

## 2 Related Work

### 2.1 The Transformative Influence of ChatGPT on Mental Health

Due to its exceptional knowledge of mental health, ChatGPT is being recognized as a promising future in providing assistance for psychological counseling (Chow et al., 2023). In the field of NLP, numerous studies have substantiated the excellent performance of ChatGPT in applications related to mental health, encompassing tasks such as stress, depression, and suicidality detection (Lamichhane, 2023; Amin et al., 2023; Kailai, 2023). Given the real-world challenges such as psychological stress and resource scarcity that human therapists may bring, ChatGPT could potentially become a significant avenue for the future of internet-based psychological counseling (Sharma et al., 2023). ChatGPT can serve as an effective tool for supplemental therapy for psychological issues, potentially addressing the mental health governance disparities in middle and low-income countries (Aminah et al., 2023; Imran et al., 2023).

### 2.2 Sentiment analysis and topic modeling of social media regarding ChatGPT

There is a burgeoning area of research in monitoring public attitudes by topic modeling and sentiment analysis from social media (Sudheesh et al., 2023). Sentiment analysis evaluates online posts as positive, neutral, or negative, offering insights into public perceptions. Topic modeling categorizes texts into themes automatically, particularly effective for analyzing social media data, like on Twitter and Reddit, commonly used to gauge public opinions on various topics.

Using this approach, numerous researchers have delved into the general public's attitudes toward ChatGPT (Sharma et al., 2023; Praveen and Vajrobol, 2023). For instance, Konchanok et al analyzed Twitter data to discern overall sentiments and topics related to ChatGPT (Koonchanok et al., 2023). Haque et al. (2022) employed Latent Dirichlet Allocation (LDA) topic modeling to pinpoint popular themes in ChatGPT-related tweets. They conducted sentiment analysis on these topics through manual labeling. Leiter et al. (2023) applied sentiment and emotion analysis to investigate public perceptions of ChatGPT, classifying tweets into 19 predetermined topics using a roBERTa-based model fine-tuned for tweet topic classification. This research provides insights into general public attitudes towards ChatGPT applications.

However, to our knowledge, there is limited research focusing on users' attitudes toward ChatGPT's application in mental health. We aim to investigate users' attitudes regarding ChatGPT's application in mental health, which can inform the ethical considerations of AI-driven mental health applications.

## 3 Methods

To uncover users' attitudes towards ChatGPT in mental health, this study collected posts related to ChatGPT in mental health from Reddit, a prominent social media. Following data cleansing, these posts underwent comprehensive sentiment analysis using the bert-base-multilingual-uncased-senbytiment model and thematic analysis utilizing the BERTopic model.

To delve deeper into the reasons behind positive and negative attitudes regarding ChatGPT in mental health, this study conducted a second round of topic analysis on the data associated with positive and negative sentiments. To understand how users' attitudes and topics evolved over time, this research statistically analyzed changes in the quantities of different emotions and topics over time. To present the analytical results more intuitively, this study employed the Matplotlib library in Python for data visualization.

The specific procedures for data collection, topic modeling, and sentiment analysis are elaborated in the following sections.

### 3.1 Data Collection And Cleaning

We utilized PRAW (Python Reddit API Wrapper), a Python library, to scrape subreddits on Reddit, a platform currently engaging approximately 430 million users. We have selected subreddits of r/ChatGPT and scrapped all the posts containing the word 'counseling', 'therapy', 'therapist', 'psychological consultant', 'emotional support', 'psychological support', ' mental health', and 'mental support'. Terms were identified by reference to papers and news reports on the application of ChatGPT to the field of mental health (Bhattacharyya et al, 2023). The collected data elements include the title, text, user name, and time of posting. Through crawlers, a total of 38875 pieces of raw data were collected in this study.

The data cleansing process involved removing duplicate entries, eliminating posts from bot users, and addressing missing data attributes. Ultimately, the database retained 18,315 records. The identification of bot users followed the methodology outlined in the referenced study (Chow et al., 2023). When a user's average time interval between consecutive posts was less than 2 hours, they were categorized as bot users, and all their posts were removed. Using this method, the study identified 7 bot users and removed 3,495 posts.

The final database is presented in the table1. It encompasses 18,315 posts contributed by 10,069 users, with an average of 1.819 posts per user. Besides, it spans from December 3, 2022, to August 23, 2023, covering a period of over 8 months. These posts are in 38 different languages, with English posts being the most prevalent, constituting 92% of the dataset. Other languages included Somali, Danish, and Pashto.

**Table 1.** Details about the Database

| Attribute | Detail |
| --- | --- |
| Date Range | 2022-12-03 to 2023-08-23 |
| Number of tweets | 18315 |
| Number of users | 10069 |
| Language counts | 38 |
| English tweets | 16861 (92%) |

### 3.2 Topic Modeling

This study employed the BERTopic algorithm for topic modeling, which is a widely acclaimed method in the field of natural language processing for topic analysis. Leveraging TF-IDF variables, clustering techniques, and transformer-based pre-trained models, BERTopic is capable of generating coherent topic representations based on textual documents (Grootendorst, 2022). These features make it highly suitable for the task of topic modeling. The workflow for topic modeling in this paper is outlined as follows.

Firstly, the raw text underwent preprocessing. Drawing from previous research, the text preprocessing encompassed several steps, including the removal of URLs, emojis, translation to English, elimination of stopwords, and lemmatization. The purpose of text preprocessing is to eliminate unrelated content, ensuring that the extracted topics closely align with the research questions. The stopwords were divided into two categories: first, the stopwords provided by NLTK ("english"), and second, custom stopwords specific to this study. Custom stopwords included common but contextually irrelevant terms within the database, such as "bot," "ai," "ChatGPT," "use," and more. Given that the research focuses on user attitudes toward "ChatGPT in mental health," keywords like "bot," "ai," "ChatGPT," "use," while prevalent, do not explicitly convey thematic meaning and may disrupt the accuracy of topic modeling results. Therefore, we removed these words. Lemmatization served the purpose of preventing different forms of the same word from being treated as separate keywords, maximizing the extraction of distinct keywords.

Next, the BERTopic technique was utilized for topic modeling. Following Koonchanok's study, the model's parameters were configured as follows: top_n_words=8, n_gram_range=(1, 2), min_topic_size=10, nr_topics=10, low_memory=True. The specific parameter values were determined through iterative refinement, based on the effectiveness of the generated topic keywords (Najmani et al., 2023). For instance, if the generated keywords are unrelated to the research question, iterations are performed. When the topic keywords hold clear meaning and relevance to the research question, the iterative process is halted, and the results are obtained.

Finally, similar topics are further merged, and irrelevant topics are filtered out. This study chose to merge topics with a cosine similarity higher than 0.7 to eliminate highly similar topics (Koonchanok et al., 2023). Ultimately, the study outputs the keywords and probability distributions of the remaining topics.

### 3.3 Sentiment Analysis

This study employed bert-base-multilingual-uncased-sentiment for sentiment analysis. The bert-base-multilingual-uncased-sentiment model is a multilingual sentiment analysis model trained on over 5,000 product reviews, achieving an accuracy rate of 95%. It is widely utilized in the field of sentiment analysis (Thang et al., 2022).

The specific process for sentiment analysis is as follows. This research utilized the tokenization provided by BERT to tokenize the text. Considering the model's text length limitation, the original text was restricted to a maximum length of 512 tokens. The encoded data was converted into a PyTorch tensor and augmented with an additional dimension using "unsqueeze(0)" to give it a batch-processing shape. Subsequently, the preprocessed data was fed into the sentiment classification model to obtain sentiment classification results. The bert-base-multilingual-uncased-sentiment model assigned sentiment scores to the text data, ranging from 0 to 4. Following prior research (Sasmaz and Tek, 2021), this paper categorized data with scores of 0 and 1 as negative, data with a score of 2 as neutral, and data with scores of 3 and 4 as positive.

To assess the model's accuracy on this dataset, scholars with a background in psychology manually annotated 100 randomly selected data as positive, negative, or neutral. By comparing the sentiment labels assigned by humans to those recognized by the machine, this study further assessed the accuracy of sentiment identification.

### 3.4 Ethical Considerations

Our data collection from Reddit adhered strictly to the platform's terms of service protocols and guidelines, ensuring the privacy and security of personal data. Our study, centered on the post level, is anticipated to have no ethical implications.

## 4 Results

### 4.1 Combined Sentiment and Topic Analysis

Combined sentiment. It is found that in discussions related to ChatGPT in mental health, negative sentiments are the most prevalent, followed by positive sentiments, with neutral sentiments being the least common. Further details are presented in Figure 1.

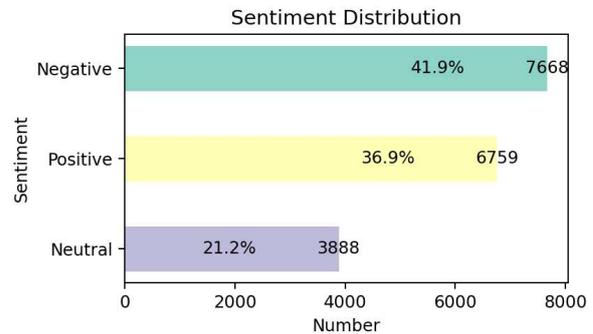

Figure 1: Distribution of positive, neutral, and negative sentiments of the full dataset predicted by BERT.

In comparison with manually annotated data, the sentiment analysis model achieved an accuracy rate of 82%, meeting the required accuracy standards. Specific instances of sentiment classification for different data categories are presented in Table 2. Among these, there were 18 instances where the human and machine judgments did not align, as illustrated partly in Table 2. In most cases of disagreement between human and machine judgment, the model categorized data with positive or negative sentiments as neutral. Based on prior research, such errors in sentiment classification are considered acceptable.

**Combined Topic Analysis.** 5 topics were extracted from discussions about ChatGPT in mental health. Detailed information about these topics, including keywords and examples, is presented in Figure 2. To facilitate understanding, researchers selected representative data from each topic as examples.

| Example | Sentiment_machine | Sentiment_manual |
|---|---|---|
| 1. ChatGPT has helped me identify and manage my emotions more than a therapist has been able to. | Positive (4) | Positive |
| 2. Mental health experts worry the high cost of healthcare is driving more people to confide in OpenAI's chatbot, which often reproduces harmful biases. | Neutral (2) | Negative |
| 3. I would threaten to just stop playing their game, and find somewhere else to invest your time and effort. | Negative (1) | Negative |

Table 2. A Comparison Between the Model's Sentiment Predictions and Manual Annotations.

The most discussed topic is 'Therapist help', with keywords including "help," "human," and "therapist." According to the examples, users discuss their experiences with ChatGPT providing assistance akin to a human therapist.

Furthermore, another extensively discussed topic is related to ChatGPT's prompt, with keywords such as "prompt," "question," and "automate." Based on the examples, it is evident that users are exchanging ideas on how to utilize prompts and questions effectively to achieve their desired outcomes.

In addition, the topic of data privacy has also garnered significant attention among users, featuring keywords like "openai," "data," "privacy," and "jailbreak." The provided examples express users' concerns regarding privacy issues associated with ChatGPT.

Moreover, there is considerable discourse regarding GPT4 and GPT3.5, with keywords including "better," "using," and "access." The examples illustrate that users are interested in the implications of ChatGPT's usage after the transition to newer versions.

The fifth topic revolves around societal well-being, with keywords encompassing "UBI" (Universal Basic Income), "capitalism," "job," and more. Using an example, users express concerns about the potential consequences of a more powerful AI in conjunction with capitalism.

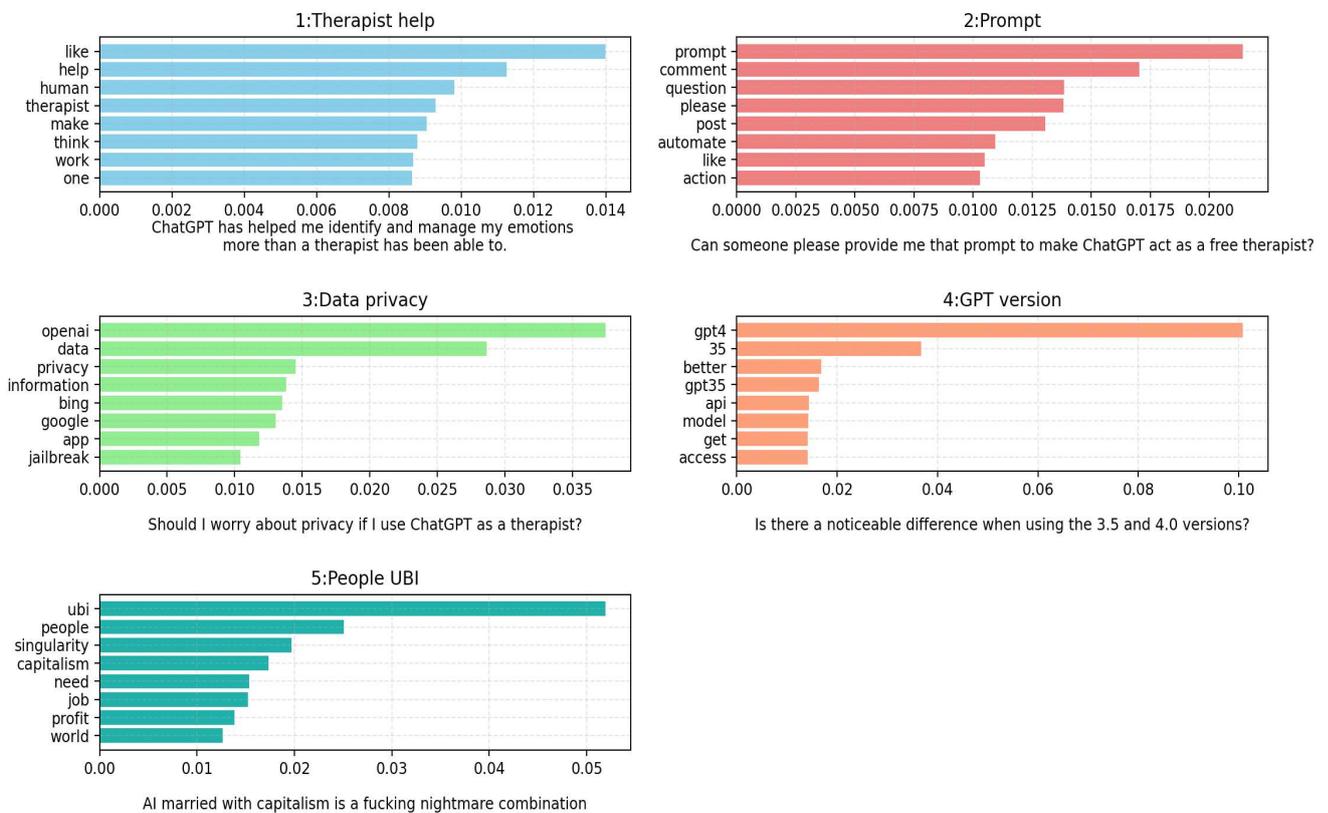

Fig. 1. Combined Topic Results and Examples Predicted by BERT.

### 4.2 Monthly sentiment and topic analysis

Monthly Sentiment Analysis. The variations in the quantity of different sentiments over time are illustrated in Figure 3, while the changes in the proportions of different sentiments within the overall sentiment are displayed in Figure 4. It can be observed that from December 2022 to March 2023, positive sentiments predominated. However, starting from April 2024, negative sentiments took the lead, with neutral sentiments consistently being the least prevalent. Figure 3 highlights that the discussions reached their peak in May, followed by a decline in the number of discussions. Notably, negative sentiments significantly surpassed positive sentiments. In the final month of data collection, August 2024, users' negative sentiment posts accounted for the highest proportion ever recorded, reaching 49.94% of all posts.

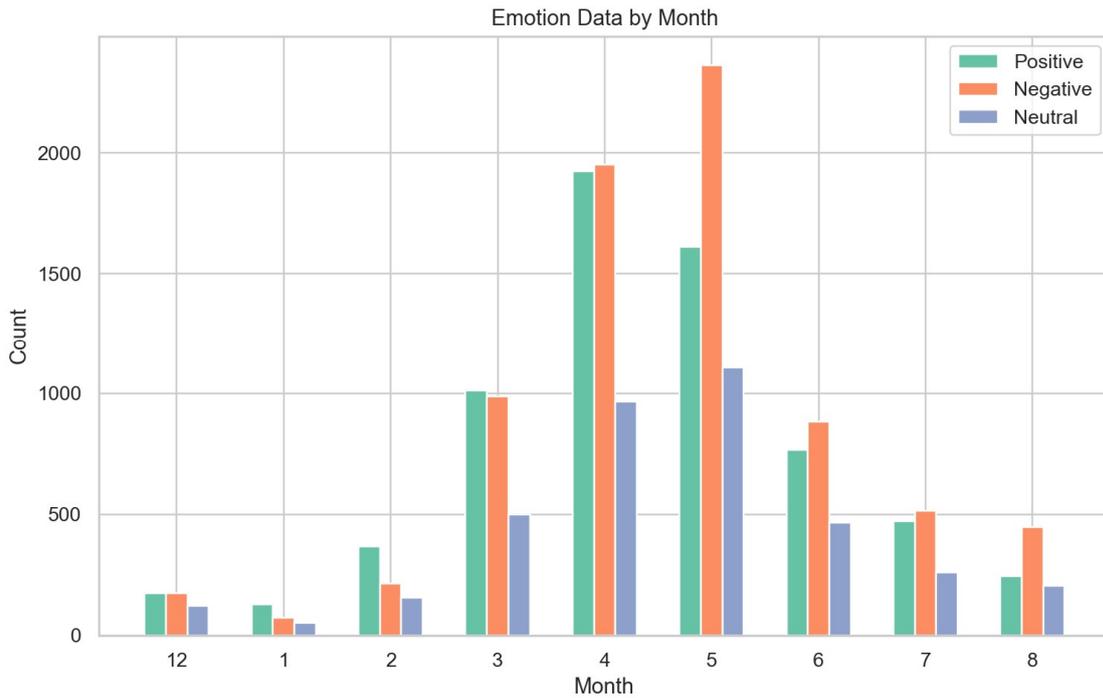

Fig. 2. Sentiment Distribution Counts by Month

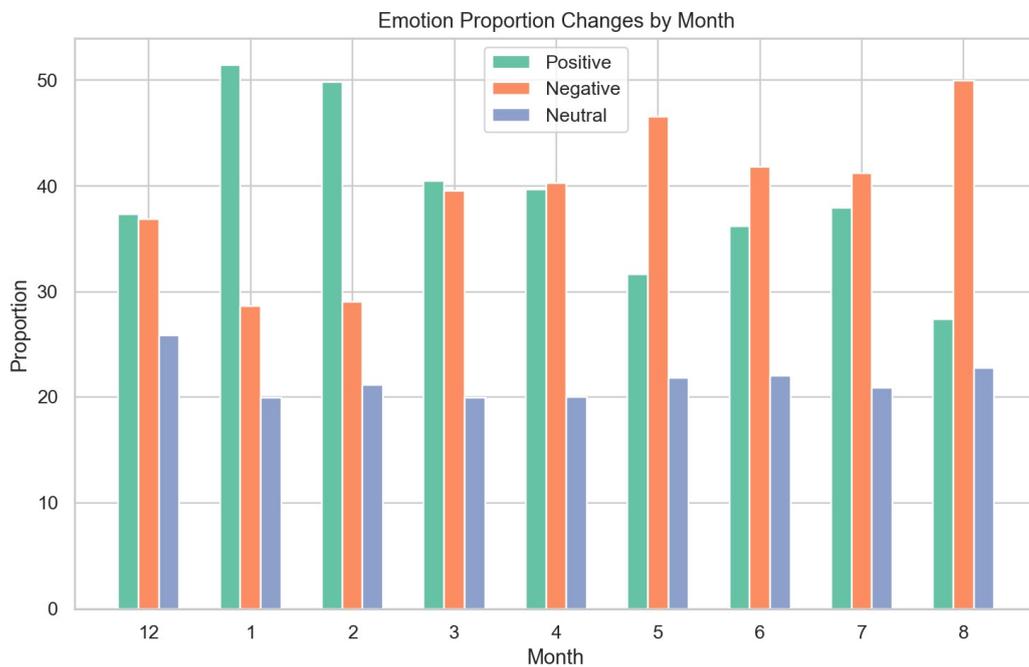

Fig. 3. Sentiment Proportion Changes by Month

**Monthly Topic Analysis.** The overall trends of different topics over time are depicted in Figure 5. It's evident that discussions related to various topics gradually increased over time, reaching their peak in May, followed by a gradual decline. The quantity of discussions for each topic is strictly correlated with its ranking. Topic 1 consistently had the highest level of discussion and reached its peak in May, subsequently decreasing. Topic 2 had the second-highest level of discussion, followed by Topic 3. Topic 4 and Topic 5 had relatively similar levels of discussion overall. However, with the release of the new GPT version in June, discussions around the new version surpassed Topic 5.

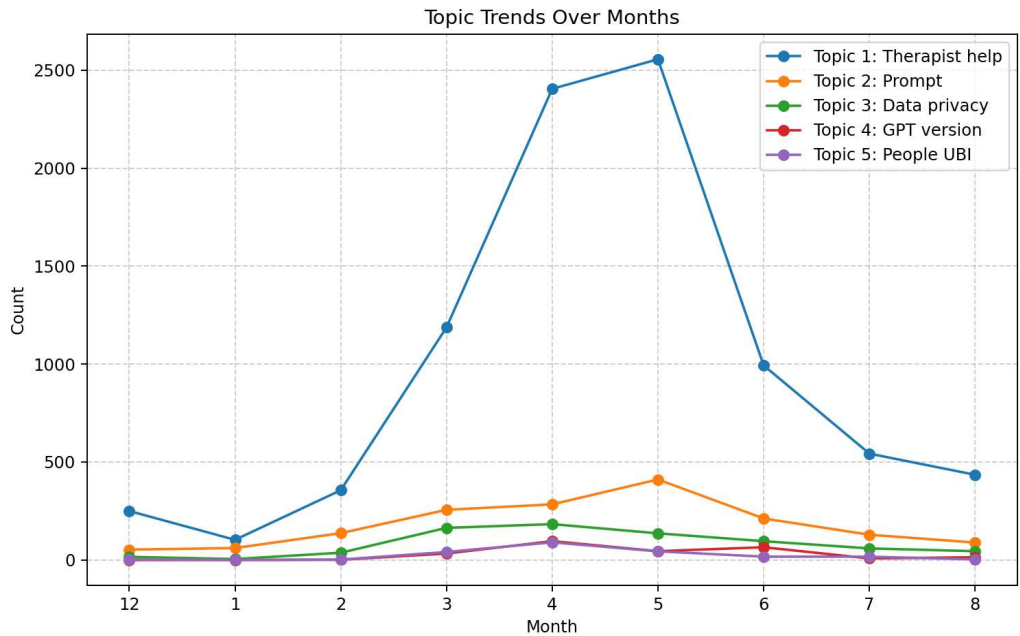

Fig. 4. Data Distribution by Different Topics over months

### 4.3 Sentiment Topic Modeling

Topic modeling results for posts with varying sentiments display a significantly heightened complexity compared to the overall dataset.

Negative Sentiment Topic. Initially, the dominant negative sentiment data underwent topic modeling, resulting in 4 topics as presented in Figure 6.

The most discussed topic revolves around "bad advice," with keywords such as "response," "language model," and dirty talk. In the examples, users pointed out that ChatGPT provides poor advice, which may potentially lead to more psychological issues.

The second topic concerns "human intelligence," with keywords including "human," "machine," "value," "stupid" and "blame." Users in the examples express the belief that AI cannot replace human intelligence.

The third topic relates to "fear," with keywords like "gun," "shooting," "existence," and "government." In the examples, users express concerns that powerful AI could be as easily accessible as firearms, leading to adverse consequences.

The final topic centers on "UBI," aligning with the fifth overarching topic in the general theme analysis, focusing on societal aspects. Keywords include "UBI," "capitalism," but also introduce "homeless" and "productivity" keywords. In the examples, individuals worry about the detrimental effects of the combination of AI and capitalism.

Positive Sentiment Topic. The topic modeling results for discussions with positive sentiments reveal 3 topics, as shown in Figure 7.

The most discussed topic is "Therapist work," with keywords like "like," "therapist," "prompt," and "time." In the examples, users express the belief that ChatGPT's assistance in mental health surpasses that of human therapists.

The second most discussed topic revolves around "love," with keywords such as "love," "future," "hope," and "thank." In the examples, users express their affection for ChatGPT's responses.

The third topic centers on "answer," encompassing terms like "interesting," "CBT" (Cognitive Behavioral Therapy), "creative answer," and "wallet." In the examples, users introduce the concept of CBT therapy prompts, a method for aiding mental health.

Neutral Sentiment Topic. The results of topic modeling for data with neutral sentiments are illustrated in Figure 8, revealing 2 topics,

including "Master information" and "GPT version."

Interestingly, discussions related to "Master information" evoke a neutral sentiment, featuring keywords such as "personal," "master," "information," "model," and "want." In the examples, users express a lack of excessive concern regarding privacy.

Furthermore, discussions surrounding GPT version updates also tend to elicit a neutral sentiment overall, with keywords like "access," "better," and "version." In the examples, users discuss the impact of version updates.

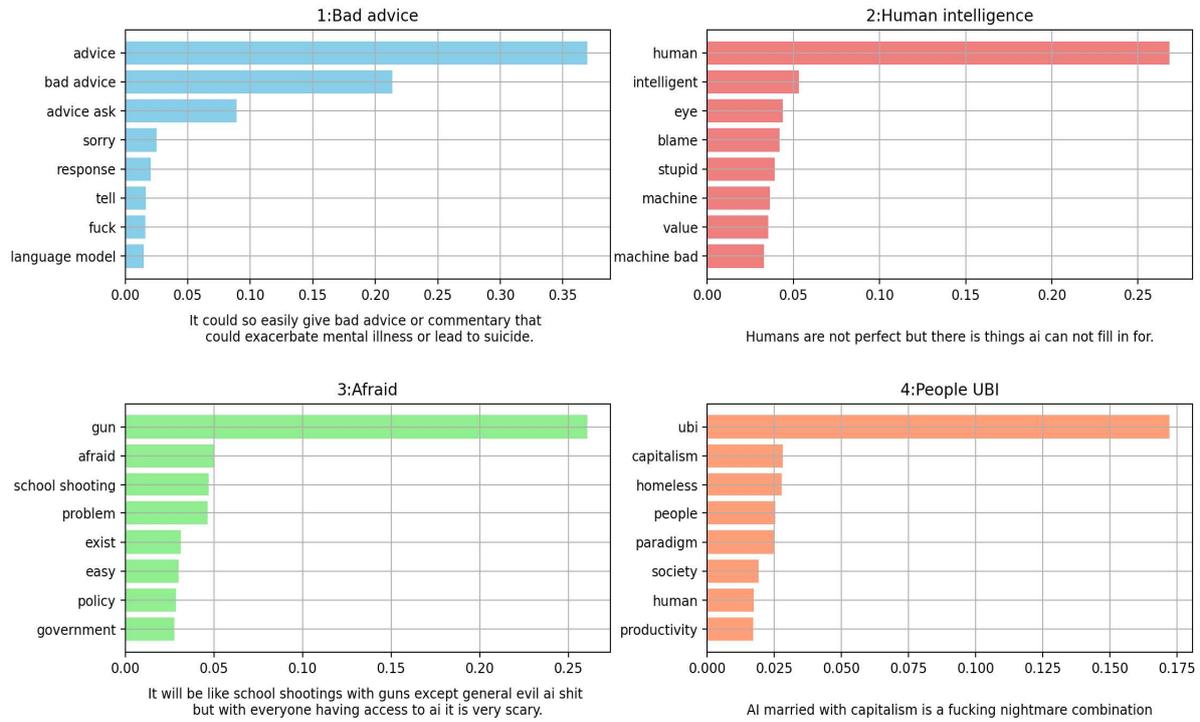

Fig. 5. Topics of negative posts

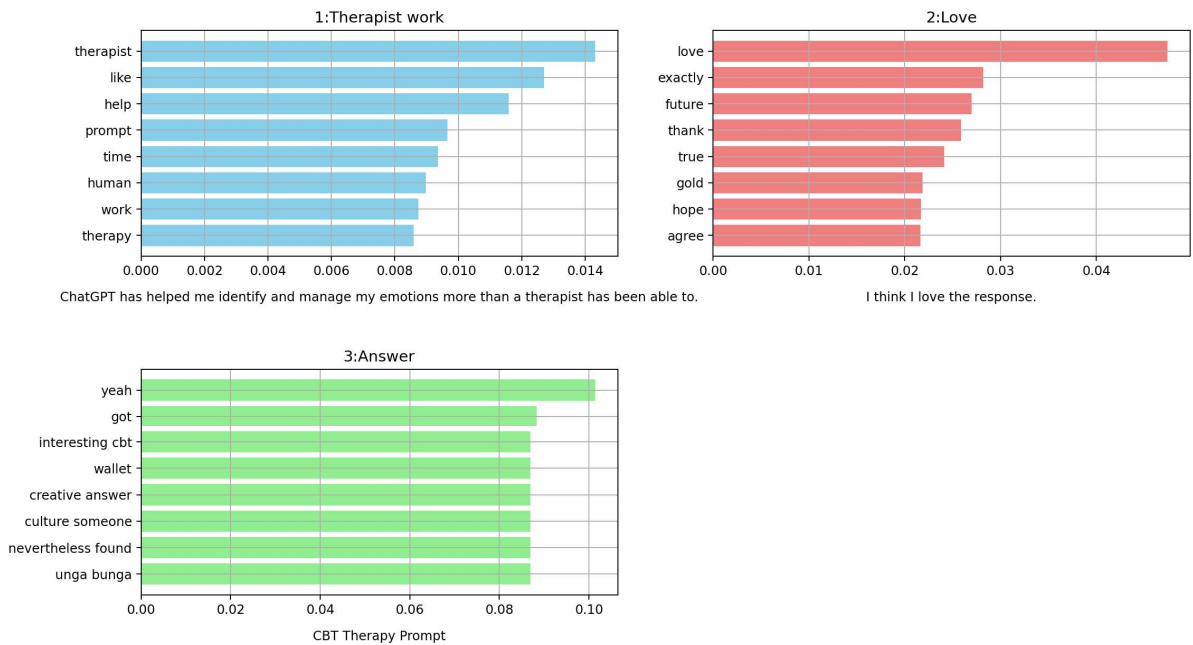

Fig. 6. Topics of positive posts

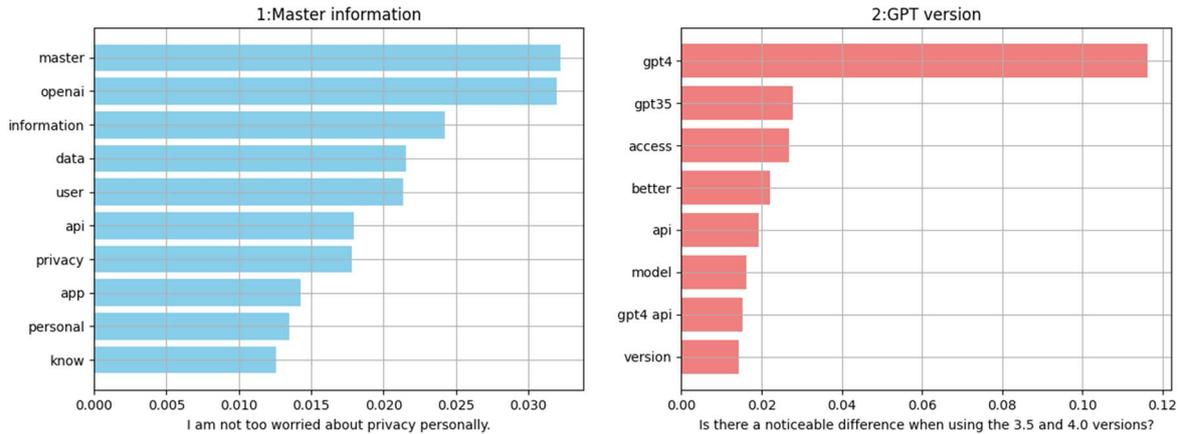

Fig. 8. Topics of negative posts

## 5 Discussion

### 5.1 Principle Findings

3 research questions were formulated to investigate user attitudes in the beginning: (1) What are the overall sentiments and topics in discussions related to ChatGPT in mental health? (2) How do the overall sentiments and topics change over time? (3) What are the topics associated with positive and negative sentiments in discussions related to ChatGPT in mental health? To address these questions, this study explored more than 18,000 posts related to ChatGPT in mental health between December 2022 and August 2023 on Reddit. We used two Bert-based models to analyze and combine sentiments and topics to identify the key public concerns towards ChatGPT application in mental health.

In response to the initial query, a higher prevalence of negative sentiments was noted compared to positive sentiments, with neutral sentiments being the least frequent. 5 topics were identified, namely "Therapist help," "Prompt," "Data privacy," "GPT version," and "People UBI."

For the second question, it was observed that the proportion of negative emotions increased over time. Additionally, discussions related to ChatGPT in mental health peaked in May and gradually declined afterward. The decrease in discussion volume may be related to a decrease in ChatGPT's traffic starting in May (Leiter et al., 2023). This could suggest that, over time, users became increasingly dissatisfied with ChatGPT in the context of mental health and gradually reduced their usage and discussions.

Regarding the third question, 4 topics were extracted from negative data: "bad advice," "human intelligence," "afraid," and "people UBI." By combining examples and keywords to further infer the content of topics, we can better understand the reasons behind users' dissatisfaction with ChatGPT in mental health. In the "bad advice" topic, it was found that users may be concerned that ChatGPT's poor advice potentially leads to more psychological problems. In the "human intelligence" topic, examples show that some users believe AI cannot replace human intelligence. Keywords like "value" and "machine bad" also express that some users prioritize human values and are against the use of AI in counseling. The "afraid" and "people UBI" topics both revolve around negative impacts of ChatGPT. In the "afraid" topic, the most common keyword is "gun". In the example, some users expressed concerns that the powerful AI could be as easily accessible as a gun and lead to disastrous consequences. In the "people UBI" topic, terms related to economic harm, such as "UBI," "capitalism," and "job," were mentioned. Combining examples, we infer that people may worry about the combination of AI and capitalism leading to increased economic exploitation.

From the data with positive sentiments, 3 topics were extracted: "Therapist work," "Love," and "ChatGPT answer." These topics can be categorized as highly positive evaluations of ChatGPT's effectiveness in mental counseling. The keyword "time" may indicate users' perception that ChatGPT responds promptly. "Wallet" might reflect users' appreciation for ChatGPT being either free or relatively inexpensive to use. Terms like "work," "human," and "like" may suggest that users view ChatGPT's work as similar to that of a human therapist.

From the data with neutral sentiments, two topics were identified: "master information" and "GPT version." The "master information" topic includes keywords like "privacy," "jailbreak," and "openai," which reflect users' concerns about privacy-related issues. This finding was unexpected, as it suggests that users have a neutral attitude toward the privacy risks associated with ChatGPT. In the examples, users indicated that they were not overly concerned about privacy issues. This may be related to the earlier discovery that users willingly disclose their personal information (Waldman, 2020).

### 5.2 Trustworthy AI in Mental Healthcare

AI in mental health proves immensely beneficial, alleviating the burdens on medical professionals and mitigating healthcare disparities, especially in resource-limited religions (Aminah et al., 2023). However, a paramount concern in AI mental health applications pertains to issues of public trust, encompassing reliability, accountability, transparency, and so on (Guo et al., 2023). Accordingly, Leveraging data gleaned from social media holds the potential for expeditious and cost-effective public sentiment and topic analysis, thereby facilitating the implementation of ethical AI development technologies and regulatory policies. For instance, users' apprehensions regarding human intelligence and inaccurate advice guide the imperative for AI healthcare development to prioritize reliability and human autonomy (Harrer, 2023). Redirecting online forum discussions, particularly those concerning ChatGPT in mental health, can conceivably contribute to fostering trustworthy AI development from the public's perspective.

### 5.3 Limitations

This study has the following limitations. First, the data collected for this study is limited to the ChatGPT-related communities on Reddit with specific keywords, and the posts were collected from December 2022 to August 2023. Therefore, all the findings in this paper are based solely on the data on Reddit during this specific time frame. It is possible that we missed other topics in online opinions.

Additionally, sentiment detection is a subjective task, which makes it difficult to explore the actual public feelings. This work is based on existing natural language processing methods which may not ensure their precise predictions reflect the actual attributes. Besides, human annotations are subject to the subjectivity of the annotators. Therefore, even with human annotation and automatic tools, it can be challenging to accurately determine the sentiment of posts.

## 6 Conclusion

Analysis of ChatGPT in mental health-related content from Reddit suggests that the sentiments expressed on social media are overall more negative than positive and negative sentiments increase with time. BERTopic modeling detected topics indicating users' different sentiments throughout the discussions. It was found that people express concerns about ChatGPT in the context of mental health, worrying about issues such as bad advice, human intelligence, and universal basic income (UBI). Overall, to our knowledge, our study is the first to offer insights into public sentiments and concerns related to ChatGPT in mental health, potentially contributing to the development of ethical AI in mental health from the users' perspective.

Future research needs to investigate this phenomenon using contemporary sentiment detection techniques, such as large language models and semantic networks, alongside an expanded dataset drawn from various social media platforms like Twitter and Facebook. Additionally, incorporating multimodal sentiment analysis considering the prevalence of videos, images, and emojis among users may contribute to a more realistic representation of population sentiment in subsequent studies.